# Chemically-distinct regions within Venus' atmosphere revealed by MESSENGER-measured $N_2$ concentrations

Patrick N. Peplowski[1*], David J. Lawrence[1], and Jack Wilson[1]

[1] Johns Hopkins University Applied Physics Laboratory, Laurel, MD 20723

[*] Corresponding Author: Patrick.Peplowski@jhuapl.edu

**Introductory Paragraph**

A defining characteristic of the planet Venus is its thick, $CO_2$-dominated atmosphere. Despite over fifty years of robotic exploration, including thirteen successful atmospheric probes and landers, our knowledge of $N_2$, the second-most-abundant compound in the atmosphere, is highly uncertain (von Zahn et al., 1983). We report the first measurement of the nitrogen content of Venus' atmosphere at altitudes between 60 and 100 km. Our result, 5.0±0.4 v% $N_2$, is significantly higher than the value of 3.5 v% $N_2$ reported for the lower atmosphere (<50 km altitude). We conclude that Venus' atmosphere contains two chemically-distinct regions, contrasting sharply with the expectation that it should be uniform across these altitudes due to turbulent mixing (e.g. *Oyama et al.,* 1980). That the lower-mass component is more concentrated at high altitudes suggests that the chemical profile of the atmosphere above 50-km altitude reflects mass segregation of $CO_2$ and $N_2$. A similar boundary between well-mixed and mass-segregated materials exists for Earth, however it is located at a substantially higher altitude of ~100 km. That Venus' upper and lower atmosphere are not in chemical equilibrium complicates efforts to use remote sensing measurements to infer the properties of the lower atmosphere and surface, a lesson that also applies to the growing field of exoplanet astronomy. The observation of periodic increases in $SO_2$ concentrations in Venus' upper atmosphere, which has been cited as evidence for active volcanic eruptions at the surface (Esposito et al., 1984), may instead be attributable to atmospheric processes that periodically inject $SO_2$ from the lower atmosphere into the upper atmosphere.

**Main Text**

Our present-day knowledge Venus' atmosphere was essentially determined in the late 1970s, when a series of descent probes and landers made in-situ measurements of the physical and chemical properties of the atmosphere. Those measurements, particularly from the Venera 11, Venera 12, and Pioneer Venus atmospheric entry probes in 1978, culminated in the development of the Venus International Reference Atmosphere (VIRA; Kliore et al., 1985). A separate evaluation of these data also led to a recommended $N_2$ concentration of 3.5±1.8 v% for altitudes <45 km (von Zahn et al., 1984). A single measurement at 55 km of ~4.5 v% $N_2$ suggested the possibility of higher $N_2$ concentrations at higher altitudes, however a measurement at altitudes >100 km was consistent with both $N_2$ values.

Our new measurement of Venus' $N_2$ content, which samples the unmeasured regime between 60 and 90 km altitude, directly resolves the question of altitude-dependent variations in $N_2$. This measurement was made using data acquired by the MESSENGER spacecraft, which performed two flybys of Venus while en route to Mercury. During the second flyby, on 5 June 2007, MESSENGER overflew the night side of Venus, with a closest approach altitude of 338 km (McAdams et al., 2007). The payload was operating and collected science data during this encounter, and those data offered an unanticipated opportunity to characterize $N_2$ concentrations via measurements of neutron emissions from the atmosphere.

MESSENGER's Neutron Spectrometer (NS) was designed to characterize the chemical composition of Mercury's surface (Goldsten et al., 2007), particularly to test the now-confirmed hypothesis that Mercury's permanently shadowed craters host large deposits of nearly pure water ice (Lawrence et al., 2013). The NS consisted of three neutron-sensitive detectors, two of which were thin plates of lithium-glass (LG) scintillator. The LG detectors measured thermal (<0.2 eV) and epithermal (0.2 eV to 0.5 MeV) neutrons from Venus, with the relative proportion being a function of the velocity of the spacecraft relative to the planet. The thermal neutrons are of interest for this study, as nitrogen acts as a "neutron poison" by absorbing thermal neutrons via the $^{14}N+n \rightarrow ^{15}N$ reaction. Consequently, atmosphere-escaping thermal neutron flux is inversely proportional to the nitrogen content of the atmosphere (Lingenfelter, Hess, and Canfield, 1962).

Neutron production is a natural consequence of galactic-cosmic-ray (GCR) bombardment of Venus' atmosphere. GCR-induced nuclear spallation reactions liberate neutrons from the atomic nuclei. As those neutrons interact with surrounding material, they lose energy via elastic and inelastic collisions, and are also subject to absorption via neutron capture reactions. The frequency and magnitude of these processes depends on the chemical composition of the atmosphere. We used the radiation transport code MCNPX (Pelowitz, 2005) to model the energy-dependent neutron flux escaping Venus' atmosphere as a function of $N_2$ content. The model included a 100-km-thick atmosphere, segmented into fifty 2-km-thick layers with uniform composition but altitude-dependent variations in the temperature and density (Seiff et al., 1986; Lodders and Fegley, 1998). We varied the $N_2$ content of all layers uniformly, from 2 v% to 7.5 vt%, in steps of 0.5 v% $N_2$, adjusting the $CO_2$ level accordingly to preserve a sum of 100 wt%. Early models that included the full chemical makeup of the atmosphere ($CO_2$ and $N_2$, plus <200 ppm of $SO_2$, Ar, $H_2O$, CO, He, Ne, and HCl) showed that inclusion of the minor elements had negligible influence on the final neutron emission spectrum, and thus they were omitted from the full set of models.

The energy-dependent shape of the atmosphere-incident GCR flux was calculated following the formalism summarized by Peplowski et al. (2019), using a GCR solar modulation parameter ($\Phi$) of 348 MV (see Methods section). The MCNPX-modeled neutron fluxes were processed using a code that transports neutrons from the top of the atmosphere to the spacecraft, accounting for gravitationally-bound neutrons on ballistic trajectories (Feldman et al., 1989), Doppler-shifting of neutron energy due to the spacecraft velocity relative to Venus (Feldman et al., 1986), and the angle- and energy-dependent neutron detection efficiency of the LG sensors (Lawrence et al., 2010). This simulation toolkit was developed and validated using Mercury flyby (Lawrence et al., 2010) and orbital (Lawrence et al., 2013, 2017; Wilson et al., 2019) datasets.

Figure 1 plots the time-series of neutron rate measurements made with the NS LG1 and LG2 detectors. Neutron count rates were derived from the raw LG spectra following a procedure detailed on the Methods section. Both detectors measured count rates that are, to first order, inversely proportional to the altitude of the spacecraft during the flyby. To second order, the count rates are sensitive to the composition of the atmosphere. The higher count rates for the LG1 detector measurements result from the fact that it was the thermal-neutron-enhanced detector during the Venus flyby, due to the velocity and orientation of the MESSENGER spacecraft. Figure 1 includes modeled neutron count rates for $N_2$ concentrations ranging from 2 to 7 v%. The half-integer v% $N_2$ models are not included in the figure for clarity. Model-to-data normalization is detailed in the Methods section.

A time-series of measured minus modeled count rates was calculated for each model nitrogen concentration using the data acquired while MESSENGER was within 1200 km of Venus' surface. The difference time series for each model was evaluated using a $\chi^2/\nu$ analysis with respect to the uncertainties of the measurements. The best-fit $N_2$ concentration is the model with the lowest $\chi^2/\nu$ value for each detector dataset (Figure 2). The one-standard-deviation uncertainty of our reported $N_2$ concentrations corresponds to those values for which $\chi^2 - \chi^2_{min} < 1$. The systematic uncertainties of the measurements, 0.3 v% $N_2$ for LG1 and 0.5 v% $N_2$ for LG2, were derived by propagating the uncertainty in the model-to-data normalization (Methods section) to the derived $N_2$ concentration values.

Our analysis yields an $N_2$ concentration and one-sigma standard deviation of 5.0±0.1 v% from the LG1 measurements (minimum $\chi^2/\nu$ = 1.43), and 5.5±0.2 v% (minimum $\chi^2/\nu$ = 1.0) for the LG2 measurements. Including the systemic errors yields 5.0±0.4 v% from LG1 and 5.5±0.7 v% from LG2. These results are consistent, so we adopt the LG1-derived value for the discussion here as LG1 was the thermal-neutron-enhanced detector during the flyby, and as a consequence it was more sensitive to $N_2$ content than the LG2 detector. This increased sensitivity is exhibited in the smaller uncertainty in the derived $N_2$ value.

Figure 3 plots in-situ $N_2$ measurements as a function of altitude, along with the new MESSENGER-NS-derived $N_2$ measurement. Our $N_2$ concentration of 5.0±0.4 v% $N_2$ is the first to sample the atmosphere at an altitude range between 60 and 90 km. The Methods section details the altitude sensitivity of the NS measurements. Our data confirm prior indications of a sharp change in $N_2$ concentrations at ~50 km altitude (Figure 3); the altitude of Venus' persistent upper cloud deck. We use this boundary to define two regions, the lower atmosphere (<50 km) and the upper atmosphere (>50 km).

Venus' lower atmosphere has the commonly cited $N_2$ concentration of 3.5 v% (von Zahn et al., 1983). On the basis of the Pioneer Venus Gas Chromatograph (PV/GC) measurements alone, the uncertainty on this value is of order 0.1 v% (Figure 3). The $N_2$ concentration of the upper atmosphere is notably higher. The PV/GC measurement is 4.60+/-0.14 v% $N_2$ at 55 km. Our measurement, 5.0±0.4 v%, is consistent with this measurement, and covers a much larger portion of the upper atmosphere. The PV Multiprobe Bus Mass Spectrometer measurement at >100-km altitude (4.3±1.3 v%) is consistent with the upper and lower atmosphere $N_2$ concentrations and is therefore not constraining.

The measurements reported in Figure 3 demonstrate that the lower- (<50-km-altitude) and upper- (>50-km-altitude) atmosphere are not mixed to chemical equilibrium. This observation rules out a well-mixed atmosphere across all altitudes (e.g. *Oyama et al.,* 1980). That the lighter-mass major constituent of Venus' atmosphere is enhanced in the higher-altitude region suggests that 50-km altitude may be the boundary at which turbulence-based compositional mixing gives way mass segregation of the two principle components ($CO_2$, mass = 44, and $N_2$, mass = 28). The NASA MSIE E-90 atmosphere model reports a similar effect on Earth (see https://ccmc.gsfc.nasa.gov/modelweb/models/msis_vitmo.php). At altitudes >100 km, the relative concentrations of lighter-mass components of Earth atmosphere (e.g. O, N, He, H) increase as a function of increasing altitude.

Venus' thick atmosphere, which has a column density >60× that of Earth's atmosphere, obscures the surface from direct observation via most remote-sensing techniques. As a consequence, observations of the upper atmosphere have been used to infer conditions at the surface. For instance, Esposito et al. (1984) attributed a sudden rise and subsequent gradual decrease of the $SO_2$ content of the upper atmosphere to an unseen, contemporaneous volcanic eruption. Similar events were subsequently observed in the 2000s by Venus Express. Although volcanic eruptions are at the surface are a possible explanation, Marcq et al. (2013) suggested that the $SO_2$ injections are instead manifestations of cyclic mixing of two chemically isolated lower and upper regions of the atmosphere. Our confirmation that two separate chemical reservoirs at low- and high altitudes do exist provides experimental evidence for that hypothesis. In this scenario, periodic increases in SO$_2$ do not reflect contemporaneous volcanic eruptions, but instead are due to infrequent mixing events between the two atmosphere regimes (Marcq et al., 2013).

Understanding Venus' lower atmosphere and surface are best accomplished via renewed in-situ exploration using atmosphere probes and landers. Our results have been cited as motivation for the planned Venera-D mission (Senske et al., 2017), which specifically proposes an in-situ investigation of altitude-dependent N$_2$ concentrations as a scientific goal. Improved knowledge of Venus also has implications for exoplanet science, as Venus is commonly used as benchmark for understanding and interpreting astronomical observations of exoplanets with thick atmospheres (e.g. Hedelt et al., 2011; Ehreneich et al., 2012). The implication of this study is a recognition of the uncertainties associated with the use of telescopic observations of an exoplanet's upper atmosphere to infer conditions in the lower atmosphere or at the surface.

**Methods**

*Neutron count rate derivation*

Both the LG1 and LG2 detectors record a 64-channel energy spectrum during each 20-s-long measurement in the timeseries. In addition to the signal of interest, which is a neutron capture peak at 4.78-MeV resulting from $^6$Li+n→$^3$H+$^4$He reactions, the spectra include a background continuum due to the interactions of GCRs and high-energy, planet-originating neutrons with the spacecraft (black curve Figure S1). The data reduction required to convert these spectra into neutron count rates involves removing this background and then summing over the channels that measure the neutron capture peak.

The neutron count rates for both the Venus and Mercury flybys were calculated using the following procedure. At high altitudes the signal is sufficiently small that the measured spectra are due only to background sources. To obtain our estimate of the background with the smallest statistical errors we sum all of the spectra taken during the flybys when the spacecraft altitude was greater than 10,000 km. Background-subtracted spectra were then produced by subtracting a scaled version of this high-altitude spectrum from each spectrum in the timeseries. The high-altitude spectrum was scaled linearly to each timeseries member such that the total counts in the background region were identical (blue curve and region in Figure S1). To calculate the total neutron count rate the background-subtracted spectra were summed over the channels around the neutron-absorption peak (red curve and region in Figure S1) and then divided by the observation period. Finally, the uncertainties in the count rate were calculated by propagating the Poisson uncertainties from the measured spectra.

*Model normalization*

Our radiation transport models account for the spectral shape of the GCR environment, but not the GCR particle fluence ($cm^{-2}\,s^{-1}$). A normalization of the model to the data is required to convert the modeled count rates to measured count rates. We performed this normalization using data acquired during MESSENGER's first Mercury flyby (M1; Lawrence et al., 2010). That flyby, which occurred on 14 January 2008, was ideal for normalization of the Venus models, as it occurred during nearly-identical GCR conditions, and with similar altitudes and attitudes relative to the flyby target.

GCR conditions are reported via a solar modulation parameter ($\Phi$), which describes the intensity and spectral shape of GCRs as they respond to changes in the solar magnetic field. The Advanced Composition Explorer (ACE) spacecraft monitors energetic particles, and $\Phi$ values are derived from those measurements (e.g. Wiedenbeck et al., 2005). $\Phi$ values derived from ACE measurements of C, O, Mg, Si, and Fe were provided to us by M. Weidenbeck (private communication). Those values were averaged to determine the $\Phi$ values on 5 June 2007 (Venus flyby; $\Phi$ = 344$\pm$30 MV) and 14 January 2008 (Mercury flyby; $\Phi$ = 324$\pm$30 MV). Within the uncertainties, these values are identical and the Venus and Mercury models were run with the Venus flyby $\Phi$ value.

Modeled count rates for the M1 flyby were derived following the same procedure developed for the Venus flyby analysis. The model included Mercury with a soil composition appropriate for the flyover geometry, which sampled equatorial regions, specifically via a closest approach over Mercury's "intermediate terrane". The soil composition, listed in Table S1 of the online supplement, was derived from x-ray and gamma-ray measured element maps of Mercury's surface (e.g. Peplowski et al., 2015; Weider et al., 2015). The Mercury flyby analysis used the same code to transport the MCNPX-modeled neutron flux at Mercury's surface to the spacecraft, accounting for Doppler enhancements of thermal neutrons, gravitationally bound neutrons, and NS detector response (Lawrence et al., 2010). As was the case for the Venus models, there parameters were calculated using the actual ephemeris of MESSENGER during the flybys.

A $\chi^2/\nu$ analysis was performed to find a single normalization parameter that provides the best-fit match between the model and the data. The best-fit normalization value is 0.55±0.03 for both LG1 and LG2 (see Figure S2 of the online supplement). The uncertainty in this normalization is propagated to our derived $N_2$ concentration as the systematic uncertainty.

A similar analysis was carried out using the Venus data and all $N_2$ models. That normalization analysis was limited to using only the high-altitude data (altitude between 1,800 and 4,000 km), where the difference between the various models was muted. Inclusion of the low-altitude data would have obscured the $N_2$-dependent effect within the model normalization. This analysis yielded a model-data normalization of 0.54, consistent with the Mercury flyby result.

*Measurement Sensitivity Altitude*

Radiation transport modeling was also used to determine the altitude-profile of the origin location of all atmosphere-escaping neutrons. Here origin is defined as the location in which a neutron last interacted with the atmosphere prior to escaping into space. This represents the location where a neutron is sampling the local composition. The model reveals a peak at an altitude of 67 km, with a full-width at half-maximum altitude range of 63 km to 75 km (Figure S3 of the online supplement) for all neutrons. These values were used as the measurement altitude range for the NS-derived $N_2$ concentrations reported in Figure 3.

**Data Availability**

The MESSENGER Neutron Spectrometer data used in this study are publicly available via NASA's Planetary Data System (https://pds-geosciences.wustl.edu/missions/messenger/index.htm). Those data include the NS LG energy spectra used to produce the measured neutron count rates shown in Figure 1 (middle and bottom panels), as well as the spacecraft ephemeris data in Figure 1 (top).

**Code Availability**

The export-controlled MCNPX radiation transport code, which formed the basis of our modeling effort, is available from https://mcnp.lanl.gov. Data processing was handled using the Interactive Data Language (IDL), with spectral processing codes following a routine detailed in Lawrence et al. (2010) and standard statistical analysis (e.g. reduced-chi-squared analyses) as detailed in the Methods section.

**Acknowledgements**

This analysis has benefited from many funding sources over a number of years. The MESSENGER mission was supported by the NASA Discovery Program under contract NAS5-97271 to the Johns Hopkins University Applied Physics Laboratory and NASW-00002 to the Carnegie Institution of Washington. Data analysis was originally supported, in part, by a MESSENGER Participating Scientist grant NNX08AN30G to DJL. Improvements to the spectral analysis and modeling codes were funded by the US Department of Energy, grant DE-SC0019343. Finally, manuscript preparation was funded by NASA's Discovery Data Analysis program, grant number NNX16AJ949 to the Johns Hopkins University Applied Physics Laboratory.

**Author Contributions**

JW produced the neutron counting rate data used in this study. DJL produced the modeled neutron count rates for all considered $N_2$ concentrations. PNP wrote the data-model comparison codes, derived the nitrogen concentration, and wrote the initial draft of the manuscript. All authors contributed to the final manuscript.

**Competing Interests**

The authors declare no competing interests.

**Additional Information**

Supplementary information is available for this paper.

Correspondence and requests for materials should be address to PNP.

**Figures**

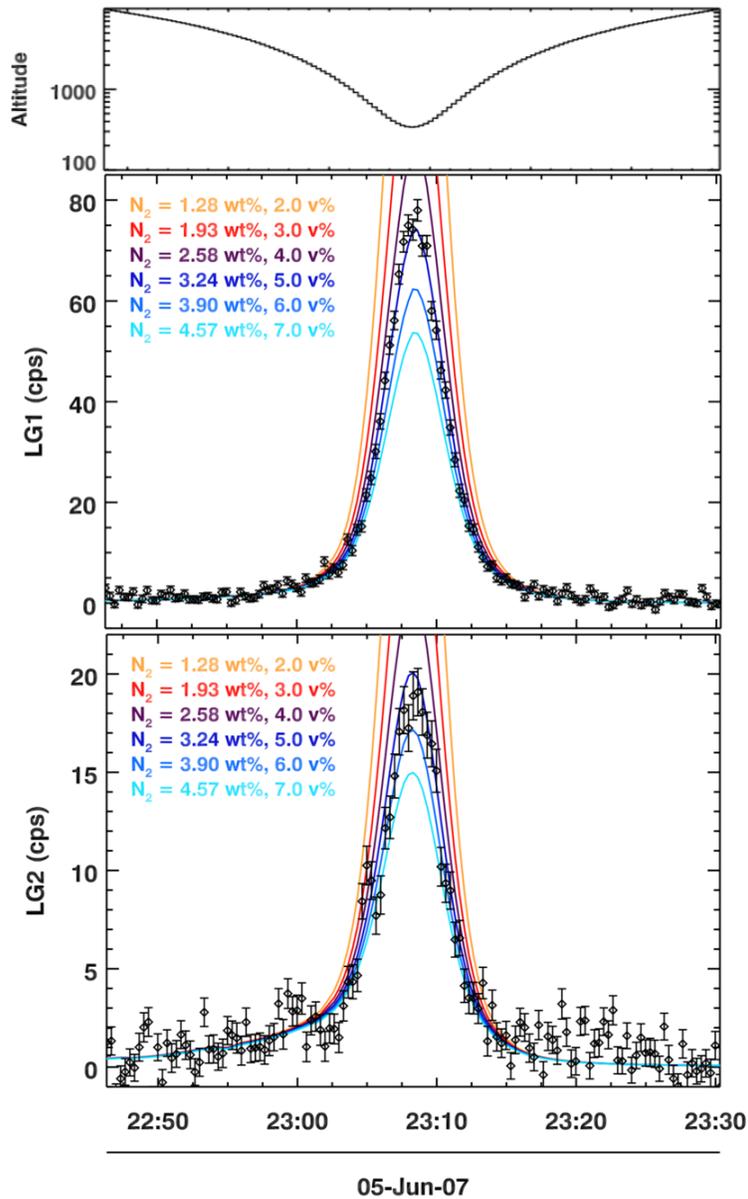

**Figure 1.** A comparison of measured and modeled neutron count rates during the 2nd MESSENGER Venus flyby. Data (black) are the neutron count rates (and 1-standard deviation statistical uncertainties) as measured by the NS "LG1" detector (middle) and "LG2" detector (bottom). Count rates increase with decreasing altitude, peaking during MESSENGER's closest approach to Venus (338-km, at 23:08 UTC; top panel). Modeled count rates are for six hypothetical $N_2$ concentrations in Venus' atmosphere. Models were normalized to the data using measurements made during MESSENGER's first flyby of Mercury.

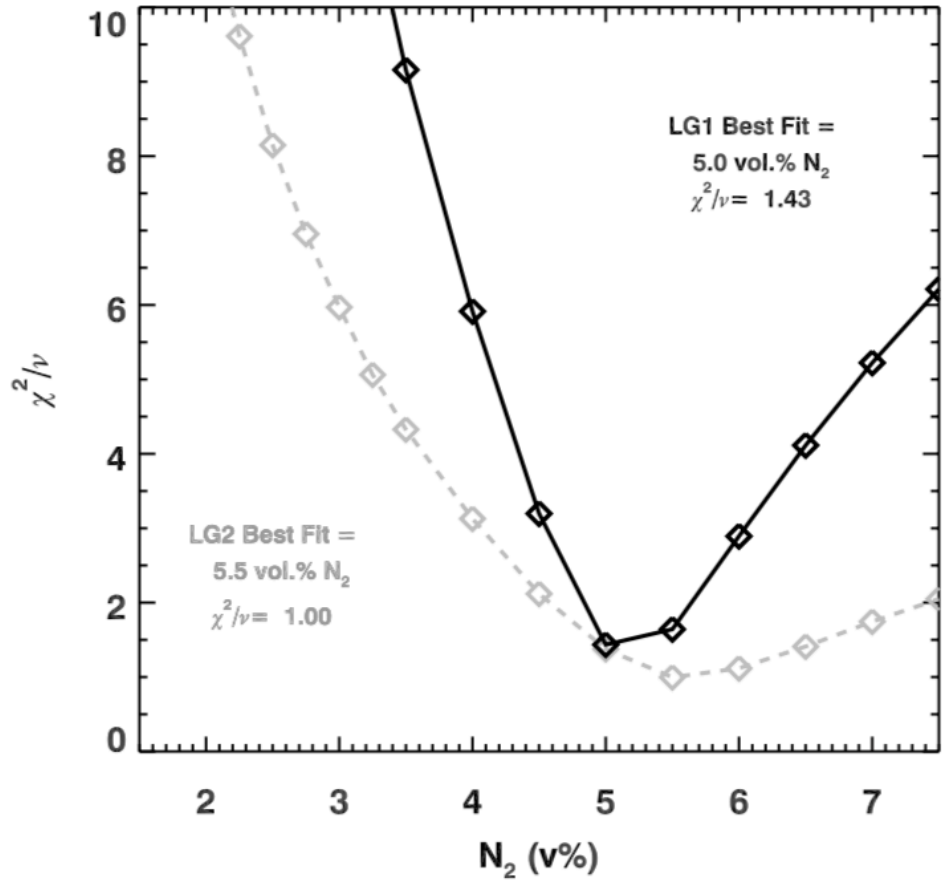

**Figure 2.** Best-fit N$_2$ values for Venus, calculated from the reduced chi-squared ($\chi^2/\nu$) values for each N$_2$ model. $\chi^2/\nu$ was calculated from the measured rates minus the modeled rates, divided by the uncertainty and summed for all measurements made at altitudes <1200 km. The N$_2$ concentration at the $\chi^2/\nu$ minimum is taken as the measured value. N$_2$ value uncertainties were calculated to encompass all N$_2$ values with $\chi^2/\nu$ <2.

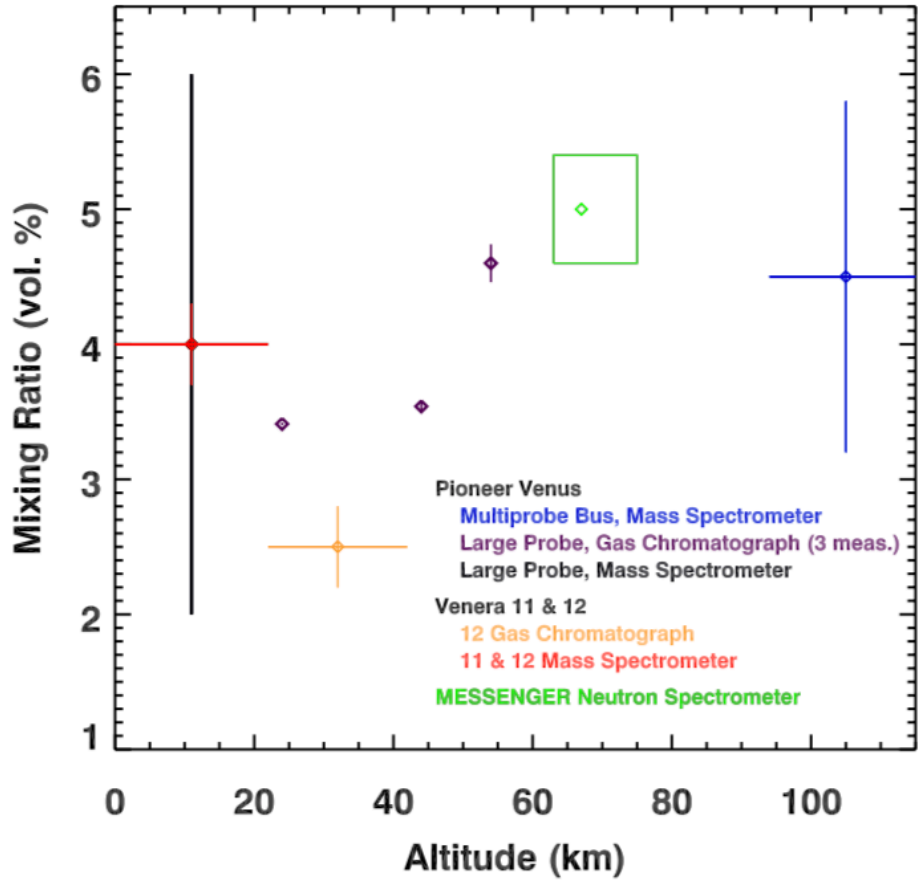

**Figure 3.** Measured $N_2$ concentrations for Venus' atmosphere, plotted as a function of the measurement altitude. The MESSENGER measurement includes the average value (diamond), and the box represents the 1-standard-deviation statistical plus systematic uncertainty range of the measurement ($N_2$ v%) and the altitude range (km) over which the measured neutrons sampled the composition of the atmosphere. The asymmetric shape of the error is due to the altitude-sensitivity profile (see Methods section). All other data are reproduced from the review of von Zahn et al. (1983).

Supplemental Material for:

**Chemically-distinct regions within Venus' atmosphere revealed by MESSENGER-measured N$_2$ concentrations**

Patrick N. Peplowski[1*], David J. Lawrence[1], and Jack Wilson[1]

[1] Johns Hopkins University Applied Physics Laboratory, Laurel, MD 20723

* Corresponding Author: Patrick.Peplowski@jhuapl.edu

*This supplement includes Figures S1-S3 and Table S1*

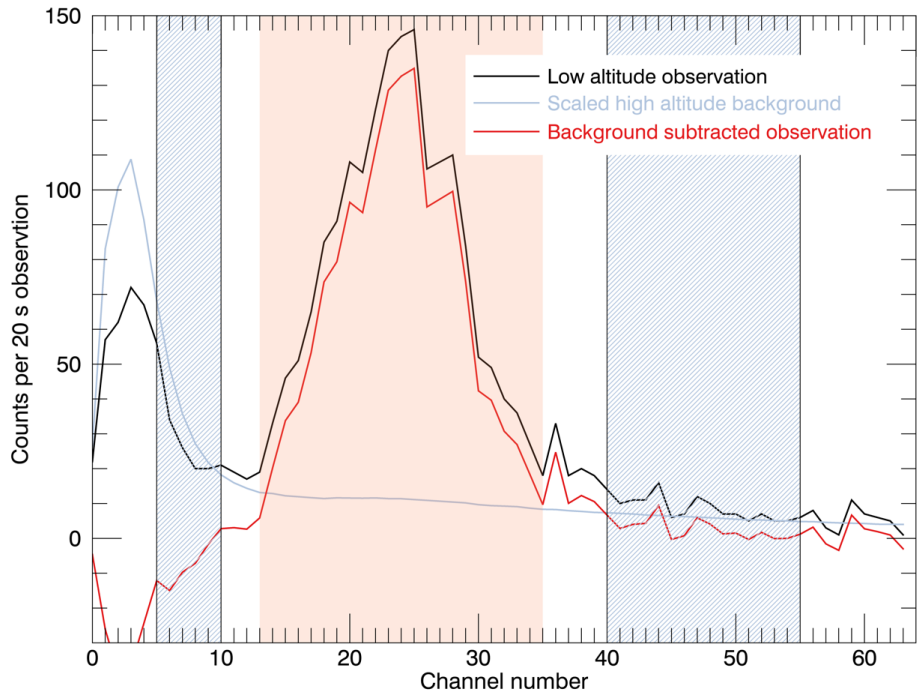

**Figure S1.** A single measured, stacked high altitude and background-subtracted spectra taken during MESSENGER's second Venus flyby. The channels highlighted in blue are those used for scaling the high-altitude spectrum to the low altitude observation. Those shown in red are the channels of the background-subtracted spectra that are summed to generate the counts in each observation period.

**Table S1.** Mercury elemental composition used for modeling of the MESSENGER M1 flyby.

| Element | Concentration (wt. fraction) |
|:---:|:---:|
| C | 0.01 |
| O | 0.3923 |
| Na | 0.0280 |
| Mg | 0.1205 |
| Al | 0.0829 |
| Si | 0.2802 |
| S | 0.0193 |
| Cl | 0.0014 |
| K | 0.0004 |
| Ca | 0.0443 |
| Ti | 0.0034 |
| Mg | 0.0011 |
| Fe | 0.0148 |

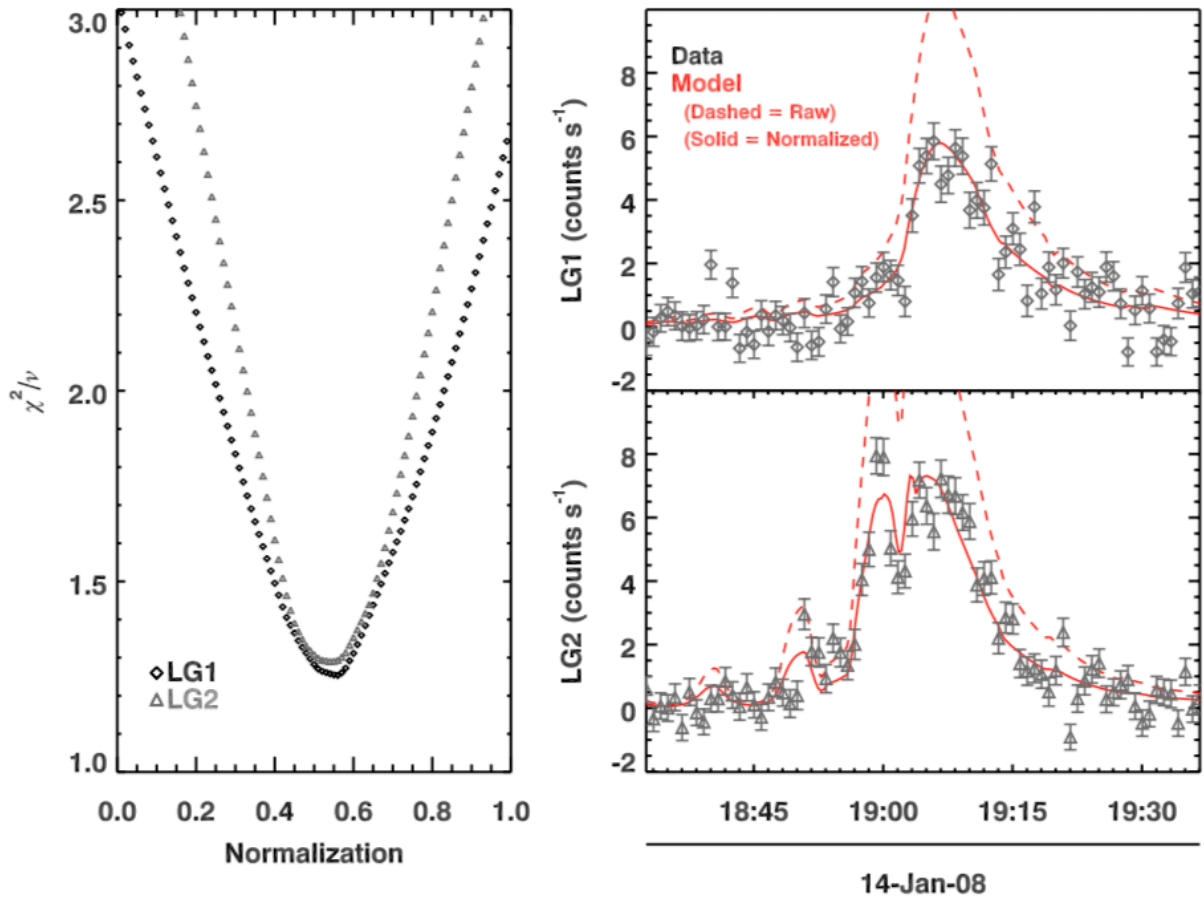

**Figure S2.** Model normalization, as determined using MESSENGER NS observations of Mercury during the first Mercury flyby. (Left) $\chi^2/\nu$ values versus model normalization. (Right) Data-to-model comparison for the LG1 (top right) and LG2 (bottom right) measurements. Model is shown before ("raw", dashed line) and after (solid line) applying the normalization.

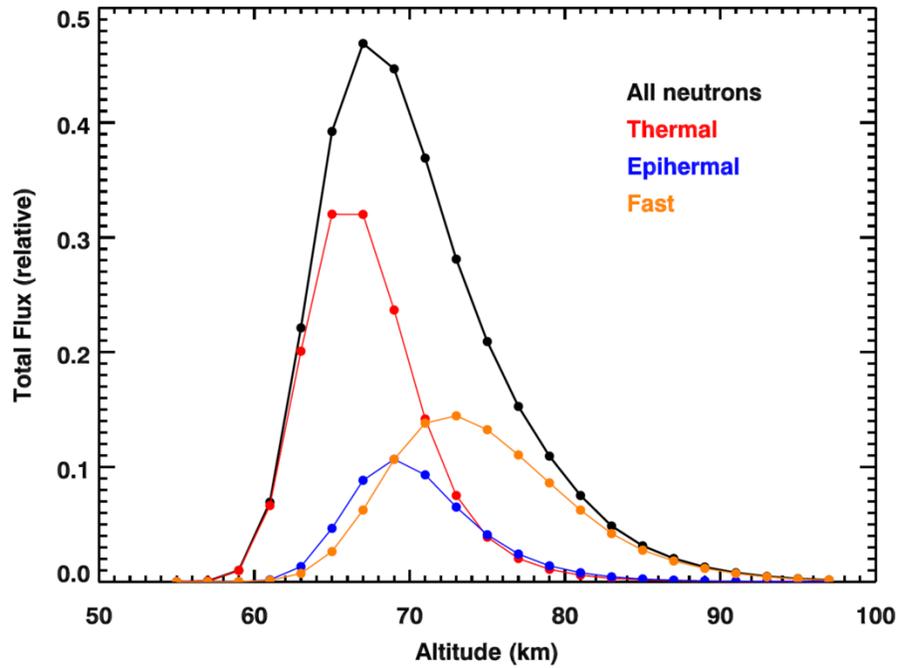

**Figure S3.** Sensitivity range of the neutron-measured $N_2$ content, as characterized using the MCNPX radiation transport model. Altitude represents the location within the atmosphere where the last interaction between a neutron and nuclei within the atmosphere occurred prior to the neutron escaping into space. This altitude corresponds to the portion of the atmosphere sampled by the neutrons.